%% file: 0_main.tex
\documentclass[11pt]{article}

\usepackage[a4paper, margin=0.9in]{geometry}

\usepackage{graphicx}
\usepackage{hyperref}
\usepackage{amsmath,amssymb}
\usepackage{siunitx}
\usepackage{authblk}
\usepackage{multicol}
\usepackage{bibentry}
\usepackage{natbib}

\usepackage{wrapfig}
\usepackage{setspace}

\usepackage{titlesec}
\titleformat*{\section}{\large\bfseries\sffamily}
\titleformat*{\subsection}{\bfseries\sffamily}

\title{\bf Exploiting light coherence in astrophysics }

\author[1]{V.~Sliusar}
\author[1]{D.~Della Volpe}
\author[1]{B.~Garcia}
\author[1]{G.~Koziol}
\author[1]{E.~Lyard}
\author[1]{N.~Produit}
\author[1]{A.~Raiola}
\author[2]{P.~Saha}
\author[2]{L.~Stanic}
\author[1]{R.~Walter}

\affil[1]{University of Geneva, Switzerland}
\affil[2]{University of Zurich, Switzerland}

\date{}

\begin{document}

\maketitle

\input{1_abstract}

\thispagestyle{empty}

\clearpage

\setcounter{page}{1}

\input{2_zbc}

\bibliographystyle{aa}
\small{\bibliography{references}}

\end{document}

%% file: 1_abstract.tex
\begin{abstract}
\noindent The Hanbury Brown–Twiss (HBT) effect, discovered in the 1950s and further developed in the 1960s, was originally used to estimate stellar angular diameters through intensity correlations measured by spatially separated detectors. Further developments started from HBT experiments to exploit quantum bunching of photons in incoherent light sources played foundational role in the development of quantum optics. 

When the two detectors in an HBT experiment are co-located, typically implemented using a beam splitter, a zero-baseline intensity correlation is obtained, which after deconvolution of the detector response function, yields the temporal component of the second-order coherence function. Unlike spatial correlations,
this function is independent of the source brightness distribution, or its size, giving direct insight into the properties of the source's emission process itself -- photon statistics. Along with photometric and spectral information, the second order coherence function can be used to constrain the emission mechanisms and discriminate between thermal, synchrotron, bremsstrahlung and stimulated emission processes. Evolution of the emission processes would likewise drive changes in the second order coherence. Light coherence information along with multi-wavelength observations, can become a complementary “messenger”, carrying internal information about the astronomical source.

\end{abstract}

%% file: 2_zbc.tex
\section*{Exploiting light coherence in astrophysics}

\begin{wrapfigure}{r}{0.4\textwidth}
\includegraphics[width=1\linewidth]{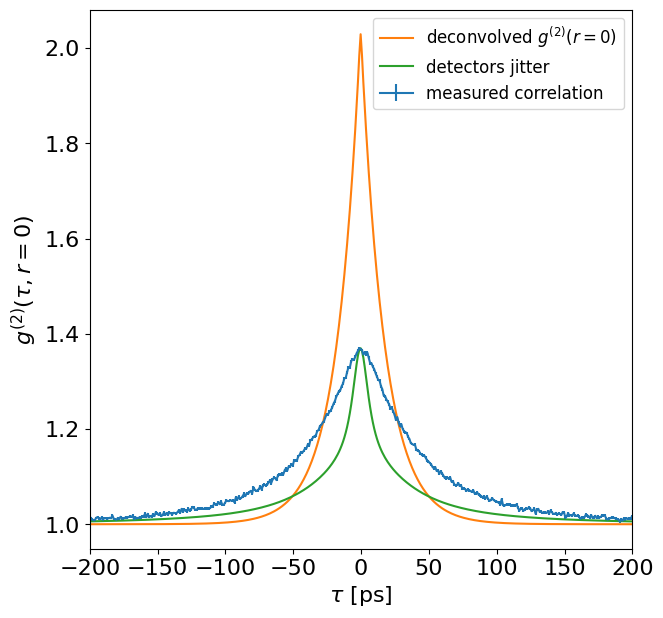} 
\caption{$g^{(2)}(0)$ (orange) for Cd 643nm line obtained by deconvolving the detector jitter (green) from the observed photon-arrival correlation (blue).}
\label{fig:zbc_hbt}
\end{wrapfigure}

The Hanbury Brown–Twiss (HBT) effect, discovered and developed in the 1950–60s, is based on measurements of the second-order coherence function $g^{(2)}(\tau,r)$. Traditionally, its spatial component is used to probe the angular size of stars or other astronomical sources, with the observed intensity correlations forming the characteristic HBT peak that depends on the separation between observing telescopes \citep{1956Natur.177...27B, brown_stellar_1967}. Beyond the spatial component, the temporal correlations of intensity fluctuations encode information about the emission process in the source itself. When applied in a zero-baseline configuration, i.e. using coincident detectors observing the same spatial mode -- the HBT effect becomes a direct probe of photon statistics, intrinsic to the source, independent of its geometry. Observed HBT effect corelation between intensities in two detectors, after deconvolution of detectors response leaves one with the second-order coherence function $g^{(2)}(0)$ (Fig.~\ref{fig:zbc_hbt}). It was previously shown \citep{sliusar_icrc2025} that $g^{(2)}(0)$ is insensitive to atmospheric disturbances, thus such measurements can provide absolute diagnostics of emission mechanisms and their evolution. $g^{(2)}(0)$ measurement requires a single telescope, without a need to resolve the source, making it particularly well-suited for compact and distant emitters such as AGNs, magnetars, and microquasars. In the context of multi-messenger astrophysics, observations of $g^{(2)}(0)$ offers a complementary “messenger”, carrying direct information on the photon statistics within the source.

\subsection*{Conceptual Basis}

In case of zero-baseline HBT effect, using Siegert relation \citep{siegert1943fluctuations, 2020AmJPh..88..831F}, for incoherent source, e.g. thermal source, the second-order correlation function can be written as $g^{(2)}(\tau=0, r=0) = 1 + |g^{(1)}(0)|^2$, where $g^{(1)}(0)$ -- first-order coherence at zero delay. For thermal or chaotic light, $g^{(1)}(0) = 1$, yielding  $g^{(2)}(0) = 2$. In contract, coherent (laser-like) or stimulated emission leads to Poissonian photon statistics with $g^{(2)}(0) = 1$, while non-classical or sub-Poissonian sources (e.g. masers under saturation or exotic quantum processes yielding photon antibunching or squeezed states) can exhibit $g^{(2)}(0) < 1$ \citep{mandel1995optical}. Therefore, the value of $g^{(2)}(0)$ directly reflects the underlying photon emission statistics. 

\subsection*{Application to Astronomical Sources}

In astrophysical contexts, measurement of $g^{(2)}(0)$ enables discrimination among several possible emission mechanisms:

\begin{itemize}
    \item \textit{Thermal Bremsstrahlung} or \textit{Synchrotron Radiation}: in general both processes are expected to produce chaotic light with $g^{(2)}(0) \sim 2$. Previous studies (e.g., \cite{1974JOSA...64.1433B}) show that spontaneous synchrotron radiation behaves as a Gaussian random (chaotic) field, implying $g^{(2)}(0) \sim 2$ under ideal single-mode detection, though in real astronomical sources the effective number of independent polarization/spatial modes (M) reduces intensity bunching from the single-mode thermal value, thus $g^{(2)}(0)=1+1/M$. More recent work \citep{2020PhRvS..23i0703L} finds that practical multi-mode, electron bunch-driven fluctuations reduce the observed photon-number variance, thus real observed $g^{(2)}(0)$ may deviate from 2. Using such measurements to confirm optical synchrotron radiation is possible for \textit{Crab Nebula}\citep{Lyutikov_Komissarov_Sironi_Porth_2018}, \textit{AGN jets} (e.g. M87 with jet blobs up to 17 mag), in star-forming H II regions, accretion environments in \textit{X-ray binaries} and \textit{pulsars}.  

    \item \textit{Coherent Emission} (e.g. in masers, pulsar radio emission): Such processes can display partial or full coherence, leading to $g^{(2)}(0)$ values approach unity \citep{2000ASPC..202..389L}. Presence of Fe II lines in \textit{$\eta$ Carinae} which are pumped by strong UV radiations leads to emergence of astrophysical lasers \citep{2005NewA...10..361J}. Another promising source may be \textit{MWC 349A} where lasing is expected during hydrogen recombination \citep{1996ApJ...470.1118S}. With VLT or ELT, using zero-baseline $g^{(2)}(0)$ measurements, we can map $\eta$ Carinae nebula (surface brightess of $\sim$8~mag/arcsec$^2$ in K \& H bands) for lasing regions.

    \item \textit{Hybrid} or \textit{Stimulated Emission Mechanisms}: In sources such as \textit{AGN jets}, \textit{fast radio bursts (FRBs)}, or \textit{magnetar flares}, a mixture of thermal and coherent components could yield intermediate correlation values $g^{(2)}(0) \in [1, 2]$, encoding the fractional contribution of each mechanism. Constraints for second order coherence will allow to study the evolution of such processes and relative contribution of each mechanism in the overall non-thermal emission. Correlate of $g^{(2)}(0)$ changes with source photometric and spectral variability would allow to better constrain physical conditions in these sources.

    \item Exotic or Quantum Regimes: In principle, if astrophysical processes were to generate squeezed or antibunched light \citep{2014PhRvA..90f3824L}, zero-baseline HBT could detect sub-Poissonian fluctuations, where $g^{(2)}(0) < 1$, although this remains experimentally challenging and requires long observations to reach high SNR values.

\end{itemize}

\subsection*{Application to Cosmology}

Strong gravitational lensing systems enable geometric measurements of the Hubble constant $H_0$ by exploiting the time delays between multiple images of a lensed variable background source, typically a quasar or a supernova. State-of-the-art projects like \textit{H0LiCOW} \citep{2020MNRAS.498.1420W} aim for percent-level precision, but remain limited by systematic uncertainties arising from the need to accurately model the lens mass distribution, the line-of-sight environment, to accurately separate intrinsic source variability from microlensing-induced one. Because gravitational lensing is strictly achromatic, it does not alter the source’s $g^{(2)}(0)$; consequently, it is insensitive to lensing-induced flux variations, thus it allows one to disentangle internal source variability from lensing effects. Monitoring the evolution of $g^{(2)}(0)$ in each lensed image separately therefore provides an alternative pathway to measure inter-image time delays, and thus to constrain $H_0$ without traditional photometric systematics. One of such sources may be RX J1131-1231 quasar with lensed images up to 16 mag.

\subsection*{Observational Considerations}

\begin{wrapfigure}[12]{r}{0.29\textwidth}
\vspace{-55pt}
\includegraphics[width=1\linewidth]{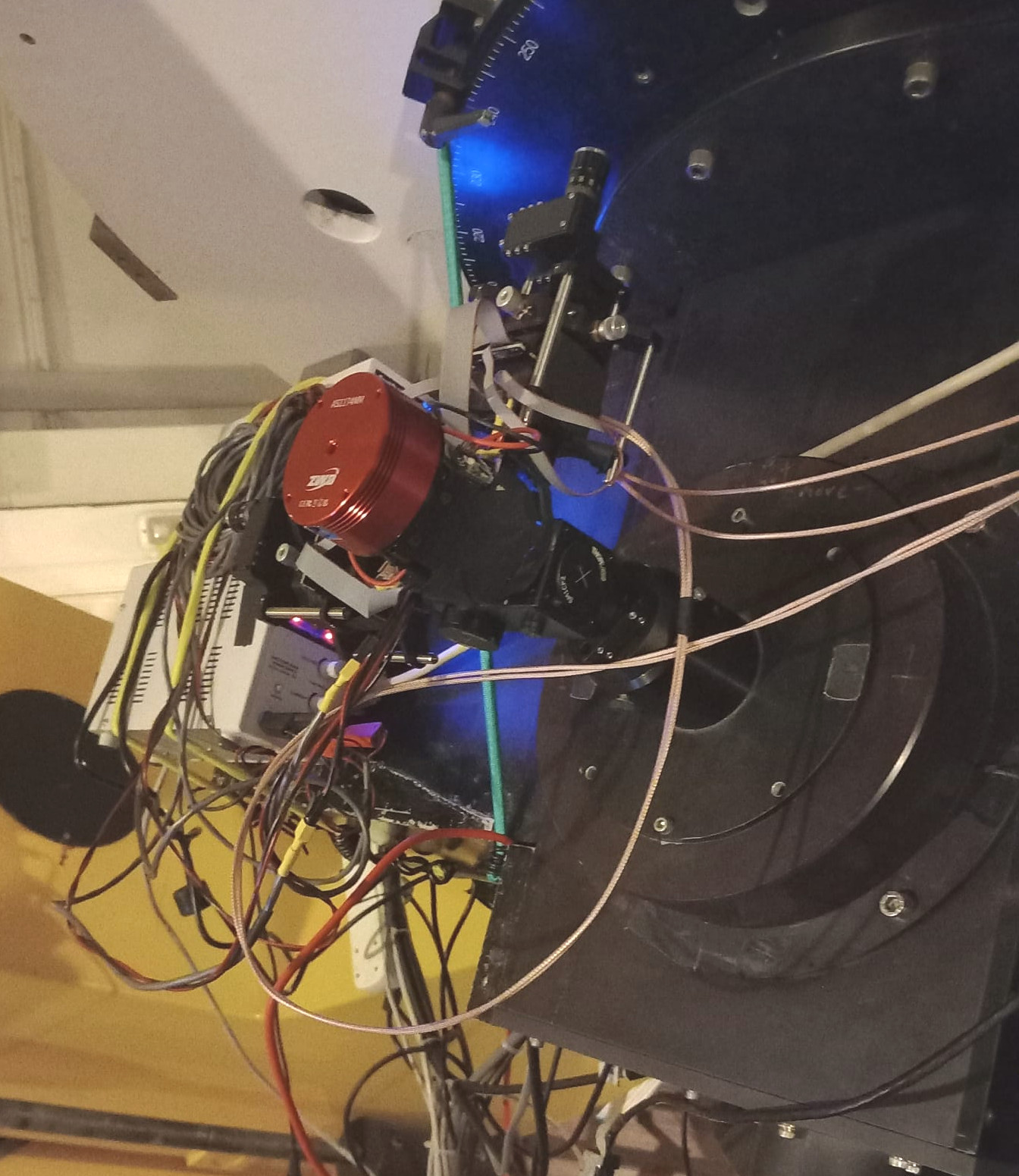} 
\caption{$g^{(2)}(0)$ instrument with 12ps jitter SPAD detectors on 1.3m telescope.}
\label{fig:zbc_setup}
\end{wrapfigure}

A zero-baseline HBT effect observations do not require spatial separation of the telescopes, but rely on high temporal resolution of detectors. Correlations must be measured over timescales comparable to the coherence time of the observed light spectral band, often requiring detectors and readout systems capable tagging photons with picoseconds accuracy. Modern single-photon avalanche diodes (SPADs) \citep{2022IJSTQ..2888216G}, superconducting nanowire single-photon detectors (SNSPDs), and fast correlators make such measurements feasible even for faint astronomical sources, like AGNs \citep{roland_icrc2025}. Modern SPAD detectors already allow measuring $g^{(2)}(0)$ (see Fig.~\ref{fig:zbc_hbt} and ~\ref{fig:zbc_setup}), such hardware can be integrated with existing telescopes, e.g. VLT, VISTA and future ones, e.g ELT.

With the VLT UT-class telescope, one can already implement a zero-baseline HBT setup by placing two fast SPAD detectors behind a narrow filter on the same telescope, and measuring photon arrival coincidences to compute $g^{(2)}(0)$. Large light collecting area of VLTs provides high photon flux, required to obtain high SNR for $g^{(2)}(0)$. For dimmer sources combination of few VLTs, measuring $g^{(2)}(0)$ independently, and then combining the signal, is also a viable option. Such VLT observations thus offers a path to test emission physics in AGN beyond spectroscopy, polarimetry or SED fitting alone. Looking ahead to the ELT, its outstanding light-collecting area (39-m class telescope) will make photon arrival statistics measurements much more feasible for even fainter, more distant, or more variable sources. 

\subsubsection*{Summary}

Zero-baseline HBT measurements of second-order coherence function shifts the focus from measuring source morphology to uncovering the quantum nature of light emitted by astrophysical objects. By analyzing photon bunching, coherence times, and deviations from classical thermal statistics, this technique can distinguish between thermal, synchrotron, coherent, and stimulated emission processes. As such, it provides a complementary diagnostic to SED model fitting, imaging, spectroscopy, and polarimetry, and opens a pathway toward applying quantum optical principles in observational astronomy. This technique also can assist measurements of $H_0$ contributing to resolution of $H_0$ tension.